\documentclass[]{interact}

\usepackage{xurl}
\usepackage[hidelinks]{hyperref}
\usepackage{listings}
\usepackage{upquote}
\usepackage{url}
\usepackage{booktabs}
\usepackage{graphicx} 
\usepackage{float}
\usepackage{epstopdf}
\usepackage[caption=false]{subfig}

\makeatletter
\def\NAT@def@citea{\def\@citea{\NAT@separator}}
\makeatother

\theoremstyle{plain}

\theoremstyle{definition}

\theoremstyle{remark}

\begin{document}

\title{Generation of Multivariate Discrete Data with Generalized Poisson, Negative Binomial and Binomial Marginal Distributions}

\author{
\name{Chak Kwong (Tommy) Cheng\textsuperscript{}\thanks{CONTACT Chak Kwong (Tommy) Cheng. Email: ccheng46@uic.edu} and Hakan Demirtas\textsuperscript{}}
\affil{\textsuperscript{}Division of Epidemiology and Biostatistics, University of
Illinois at Chicago, 1603 W. Taylor St., Chicago, IL 60612}
}

\maketitle
\begin{abstract}
The analysis of multivariate discrete data is crucial in various scientific research areas, such as epidemiology, the social sciences, genomics, and environmental studies. As the availability of such data increases, developing robust analytical and data generation tools is necessary to understand the relationships among variables. This paper builds upon previous work on data generation frameworks for multivariate ordinal data with a prespecified correlation matrix. The proposed algorithm generates multivariate discrete data from marginal distributions that follow the generalized Poisson, negative binomial, and binomial distributions. A step-by-step algorithm is provided, and its performance is illustrated in four simulated data scenarios and three real-data scenarios. This technique has the potential to be applied in a wide range of settings involving the generation of correlated discrete data.
\end{abstract}

\begin{keywords}
Collapsing, Generalized Poisson Distribution, Negative Binomial Distribution, Binomial Distribution, Random Number Generation, Correlated Discrete Data
\end{keywords}

\section{Introduction}
  Stochastic simulation plays an important role in statistical and scientific research. By generating synthetic data, researchers can simulate complex real-world data related to their studies, assess the performance of statistical methods under various conditions, and evaluate the properties of statistical models, including the bias and consistency of estimators, as well as sensitivity and specificity. Data featuring multivariate discrete responses are very common in many research areas, including medicine, the natural sciences, finance, and the social sciences. For example, in a clinical trial study, Thall and Vail (1990) and Fotouhi (2008) analyzed the dataset from a longitudinal study of epileptic patients. Patients were randomized into a treatment group with progabide or a placebo group. The number of seizure occurrences among the 59 epileptic patients was reported over four successive biweekly clinic visits. In an epidemiological research study, Martin et al. (2018) investigated the association between treatment and soil-transmitted helminth infection on microbial composition at the phylum level. The microbial counts of six bacterial phyla were analyzed to examine their relationship with helminth infection status. Soneson and Delorenzi (2013) performed an exhaustive comparison of differential expression analysis methods using RNA-sequencing data to identify genes that exhibit differences in expression levels.

Several methods for simulating correlated count data appeared in the literature. Yahav and Shmueli (2012) developed an algorithm for generating multivariate Poisson distributions using an improved version of the NORTA method (NORmal To Anything). Li et al. (2021) further proposed an adaptive approach to the NORTA method for generating data that follow multivariate generalized Poisson distributions. However, both approaches depend on a single distribution type, which lacks flexibility in distributional choices. Madsen and Dalthorp (2007) proposed two algorithms for simulating correlated Poisson data. The lognormal-Poisson hierarchy method is computationally efficient, but it is limited to low correlations with small means of the marginal variables. It also requires Poisson variables to be over-dispersed. In contrast, the overlapping sums method can handle under-dispersed Poisson variables. Still, it is restricted to positive correlations and becomes too complex to implement when the number of variates exceeds three. Minhajuddin and Juarez (2013) proposed a mixture approach to generate multivariate count data, which requires a joint distribution consisting of a latent variable and the product of the conditional distributions of the individual count variables given that latent variable. However, it can only generate data with positive exchangeable structures. 

To the best of our knowledge, recent research on generating multivariate discrete data remains relatively limited. The proposed mechanism in this paper builds upon the methodology developed by Demirtas (2006) for generating correlated ordinal data. Our method provides a comprehensive framework for generating three major multivariate count distributions: the generalized Poisson, negative binomial, and binomial distributions. Another highlight of our approach is that, unlike methods that heavily depend on a joint distribution model, our method requires only the specification of marginal distributions and a feasible correlation matrix. Additionally, our approach can accommodate various correlation structures, including both positive and negative correlations, which closely mimic the complexity of real-world data.

The remainder of the paper is organized as follows. Section 2 introduces the statistical properties of the distributions considered and provides a detailed description of the proposed algorithm. Section 3 evaluates the methodology through simulation studies, focusing on the accuracy and precision of parameter estimates. Section 4 introduces a computational tool in the form of an R package and addresses implementation-related considerations. Section 5 presents three real-data applications to further assess the performance of the method. Finally, Section 6 concludes with a discussion and outlines potential directions for future research.

\section{Multivariate Discrete Data Generation}

Below are the univariate discrete distributions that form a basis for the multivariate versions in our data generation algorithm.

\subsection{Univariate Generalized Poisson distribution }\label{class}

Using the notation in Consul (1989) and Demirtas (2017), let $X$ be a generalized Poisson variable with rate parameter $\theta$ and dispersion parameter $\lambda$, its probability mass function is described as follows:

\[
P(X = x) =
\begin{cases}
\displaystyle 
\frac{\theta(\theta + \lambda x)^{x-1} e^{-\theta - \lambda x}}{x!}, & \text{for } x = 0,1,2,\dots \\
0, & \text{for } x > m \text{ if } \lambda < 0
\end{cases}
\]

and zero otherwise, where $\theta > 0$, $\max(-1,-\theta/m) \le \lambda < 1$, and $m(\ge 4)$ is the largest positive integer for which $\theta + m\lambda > 0$ when $\lambda < 0$. The case $\lambda = 0$ corresponds to the regular Poisson distribution. When $\lambda > 0$ or $\lambda < 0$, the distribution exhibits over-dispersion or under-dispersion relative to the regular Poisson distribution, respectively. The expected value is given by $E(X)=\theta(1-\lambda)^{-1}$ and the variance is $\mathrm{Var}(X)=\theta(1-\lambda)^{-3}$.

\subsection{Univariate Negative Binomial Distribution}\label{class}
Let $Y$ be a negative binomial random variable, where $y$ is the number of failures before the $r^{\text{th}}$ success, $r$ is the number of successful trials, and $p$ is the probability of success. The probability mass function is given by

\[
P(Y = y) = \binom{r + y - 1}{y} p^{r}(1-p)^{y}, \quad y = 0,1,2,\dots
\]

with the expected value $E(Y) = \dfrac{r(1-p)}{p}$ and variance $\mathrm{Var}(Y) = \dfrac{r(1-p)}{p^{2}}$.

\subsection{Univariate Negative Binomial Distribution}\label{class}

Let $V$ be a binomial random variable, where $v$ is the number of successes, $n$ is the number of trials, and $p$ is the probability of success for each trial. The probability mass function is given by

\[
P(V = v) = \binom{n}{v} p^{v}(1-p)^{n-v}, \quad v = 0,1,2,\dots, n,
\]

with the expected value $E(V) = np$ and variance $\mathrm{Var}(V) = np(1-p)$.

\subsection{Correlation Boundaries}\label{class}
When simulating correlated random variables, it is important to determine the feasible range of correlations for each pair of variables to ensure that the desired correlations fall within these limits. Hoeffding (1994) and Fréchet (1951) demonstrated that the lower and upper bounds for the correlation between two random variables depend on their marginal distributions. Let $X$ and $Y$ be real-valued random variables with cumulative distribution functions (CDFs) $F$ and $G$, respectively. Denote $\Pi = \Pi(F,G)$ as the class of all bivariate CDFs $H$ on $\mathbb{R}^2$ with marginals $F$ and $G$. Hoeffding and Fréchet established the existence of two bivariate CDFs $H_L(x,y)$ and $H_U(x,y)$ within $\Pi$, called the lower and upper bounds, which have minimum and maximum attainable correlation. For all $(x,y)\in\mathbb{R}^2$,
\[
H_L(x,y)=\max\{F(x)+G(y)-1,0\}
\]
and
\[
H_U(x,y)=\min\{F(x),G(y)\}.
\]
For any $H\in\Pi(F,G)$ and all $(x,y)\in\mathbb{R}^2$, $H_L(x,y)\le H(x,y)\le H_U(x,y)$. Let $\delta_L,\delta_U,\delta$ be the correlation coefficients for $H_L,H_U,H$ respectively, then $\delta_L\le \delta \le \delta_U$. If $U$ is uniform in $[0,1]$, then the correlation bounds are given by
\[
\left\{\mathrm{Corr}\left(F^{-1}(U),G^{-1}(1-U)\right),\;
      \mathrm{Corr}\left(F^{-1}(U),G^{-1}(U)\right)\right\}.
\]
Further, the approximation of these lower and upper correlation bounds can be made using Demirtas and Hedeker’s (2011) Generate, Sort, and Correlate (GSC) algorithm. All correlations in this paper refer to Pearson product-moment correlations.

\subsection{Generation of Multivariate Binary Variables}\label{class}
Emrich and Piedmonte (1991) proposed a latent normal variable approach to generate multivariate binary data. Let $Y_1,\dots,Y_J$ denote the outcomes for binary variables, let $p_j$ denote the marginal expectation of $Y_j$ for $j=1,\dots,J$ and $\delta_{jk}=\mathrm{corr}(Y_j,Y_k)$ denote the pairwise Pearson correlation of $Y_j,Y_k$ for $j=1,\dots,J-1$ and $k=2,\dots,J$. Further, let $\Phi[x_1,x_2;\rho]$ be the standard bivariate normal cumulative distribution function with correlation coefficient $\rho$ as

\[
\Phi[x_1,x_2;\rho] = \int_{-\infty}^{x_1}\int_{-\infty}^{x_2} f(z_1,z_2;\rho)\,dz_1\,dz_2,
\]

where

\[
f(z_1,z_2;\rho)
=
\left[2\pi(1-\rho^2)^{1/2}\right]^{-1}
\exp\!\left(
-\frac{z_1^2-2\rho z_1 z_2+z_2^2}{2(1-\rho^2)}
\right).
\]

Solving the equation

\[
\Phi\!\left[z(p_j),z(p_k);\rho_{jk}\right]
=
\delta_{jk}\left[p_j(1-p_j)p_k(1-p_k)\right]^{1/2}
+
p_j p_k
\]

for $\rho_{jk}$ yields the intermediate correlation of the standard normal variables to generate the correlated binary outcomes. Here $z(p)$ is the $p^{\text{th}}$ quantile of the standard normal distribution.

Next, generate a $J$-dimensional multivariate normal random variable
$Z=(Z_1,\dots,Z_J)$ with mean $\mathbf{0}$ and correlation matrix
$\Sigma=(\rho_{jk})$, which is positive semidefinite. After that, for
$j=1,\dots,J$, assign $Y_j=1$ if $Z_j\le z(p_j)$ and $Y_j=0$ otherwise.

\subsection{Algorithm for Generating Multivariate Generalized Poisson, Negative Binomial and Binomial Data}\label{class}
Our methodology of generating multivariate generalized Poisson, negative binomial, and binomial data is an extension of the approach for generating multivariate ordinal data proposed by Demirtas (2006). Since the generalized Poisson and negative binomial distributions have unbounded support, we first define a pseudo upper limit for each of these marginal distributions by truncating values with negligible point probabilities. Specifically, we define the pseudo upper limit $K$ as the smallest integer in the upper region of the support such that all larger values have point probabilities less than or equal to $1\times 10^{-10}$.

\vspace{1em}
Let $Y_1,Y_2,\dots,Y_J$ be the $J$ generalized Poisson / negative binomial / binomial variables with $P(Y_j=k)=p_{jk}$, $N$ be the total number of observations, and let the pairwise Pearson correlation be $\mathrm{corr}(Y_i,Y_j)=\delta_{ij}^*$ for $i\neq j$, where $i=1,\dots,J$, $j=1,\dots,J$ and $k=0,\dots,K$. The algorithm is described as follows:

\begin{enumerate}
\item Analogous to the multivariate ordinal data generation approach, treat each discrete outcome in generalized Poisson, negative binomial, and binomial random variables as a ``category.'' First, collapse the univariate generalized Poisson/negative binomial/binomial outcome into a binary variable. Let $M$ be the median of the marginal discrete variable. We then assign outcomes smaller than $M$ as 0 and outcomes greater than $M$ as 1. The median outcome is then assigned to the binary category that makes the expectation of the binary variable closer to or equal to 0.5, as binary variables perform best in the neighborhood of 0.5. Denote the newly formed binary variable as $Y_j^b$ with $p_j^b = E[Y_j^b]$.

\item Compute the corresponding binary correlation $\delta_{ij}^b$ via simulation. We use the following iterative procedure that uses the property $|\delta_{ij}^b| \ge |\delta_{ij}^*|$ (Demirtas 2006) for all $i$ and $j$ where $i \ne j$ under the large sample assumption.

\begin{enumerate}
\renewcommand{\labelenumii}{(\alph{enumii})}

    \item Set the starting value of $\delta_{ij}^b$ to be $\delta_{ij}^*$.

    \item Generate the $N \times J$ matrix of multivariate binary data ($J=$ number of variates) using the Emrich and Piedmonte (1991) approach with a large number of samples (e.g., $N_{\text{binary}} = 100{,}000$). This approach computes the intermediate correlation matrix $\Sigma^*$ for a multivariate normal distribution.

    \item Convert the binary data generated in (b) to the generalized Poisson / negative binomial / binomial integer scale with the original proportion $p_{jk}/(1-p_j^b)$ and $p_{jk}/p_j^b$ for binary categories 0 and 1, respectively. Then, compute a pairwise correlation, denoted as $\delta_{ij}^{*c}$. If the difference between $\delta_{ij}^*$ and $\delta_{ij}^{*c}$ is within the tolerance range (e.g., 0.001), go to Step 3; otherwise, continue to the next step (d).

    \item Find the absolute difference between $\delta_{ij}^*$ and $\delta_{ij}^{*c}$, increase $\delta_{ij}^b$ by a small fraction of the difference, and repeat Steps 2(b) to 2(c).
\end{enumerate}

\item Repeat Step 2 for each pair of variables $(J \times (J-1))/2$ times and assemble all the pairs into an overall binary correlation matrix $\Sigma$ whose off-diagonal elements are the final $\delta_{ij}^b$ computed in Step 2. A positive definite and valid $\Sigma$ is not always guaranteed. Then compute the nearest positive definite matrix of $\Sigma$ using the methods proposed by Higham (2002) using the \texttt{nearPD} function (Bates et al.\ 2025) in the R package \texttt{Matrix}.

\item Generate binary data using $p_j$ and $\Sigma$ using the Emrich and Piedmonte (1991) approach.

\item Convert the binary data generated in Step 4 to generalized Poisson / negative binomial / binomial scale using the proportion $p_{jk}/(1-p_j^b)$ and $p_{jk}/p_j^b$ for binary categories 0 and 1, respectively for any desired sample size.
\end{enumerate}

\section{Method Evaluation via Simulation Study}

We consider four simulation scenarios to evaluate the proposed method for each of the multivariate discrete distributions. For each scenario, we generate 1,000 datasets under both small ($N=200$) and large ($N=2000$) sample sizes. To evaluate performance, we report the measures of accuracy and precision in each case study. We present the true values (TV), the average estimates (AE), the standard deviation of estimates (SD), the relative bias (RB) of the estimates using $|(E(\hat{\theta})-\theta)/\theta|\times 100\%$, and the standardized bias (SB) using $|(E(\hat{\theta})-\theta)/SD(\hat{\theta})|\times 100\%$, the root mean square error (RMSE) $\sqrt{E(\hat{\theta}-\theta)^2}$, and the coverage rate (CR) across 1,000 simulated datasets. We use the method of moments to compute the average estimates of the parameters from the simulated datasets. Based on the evaluation criteria in Demirtas (2004) and Amatya and Demirtas (2015), $RB<5\%$, $SB<50\%$, and $CR>90\%$ are within acceptable limits. All values of AE, SD, RB, SB, and RMSE are computed from the simulation outputs and subsequently rounded to four decimal places for reporting.

\subsection{Multivariate Generalized Poisson Distribution }
First, we consider a multivariate discrete distribution consisting of five generalized Poisson variables with various rate and dispersion parameters, for
\[
\boldsymbol{\theta} = (5.14,\,10.67,\,30.38,\,50.02,\,2)
\]
and
\[
\boldsymbol{\lambda} = (0.6445,\,0.1420,\,-0.1378,\,-0.0499,\,0.365).
\]
This specification covers a reasonably broad range of the distribution parameters. Further, we specify the unstructured correlation matrix $\Sigma$ as
\[
\Sigma =
\begin{pmatrix}
1      & 0.0644 & 0.1041 & -0.0658 &  0.2619 \\
0.0644 & 1      & 0.1008 &  0.1246 & -0.0122 \\
0.1041 & 0.1008 & 1      &  0.0867 &  0.1724 \\
-0.0658& 0.1246 & 0.0867 &  1      &  0.0452 \\
0.2619 & -0.0122& 0.1724 &  0.0452 &  1
\end{pmatrix}
\]

\begin{table}[H]
\centering
\caption{Results from the generalized Poisson distribution simulation in a small sample case ($N = 200$).}
\label{tab:gp_small}
\scriptsize
\setlength{\tabcolsep}{5pt}
\renewcommand{\arraystretch}{0.92}

\begin{tabular}{lrrrrrrr}
\toprule
      & TV    & AE     & SD     & RB     & SB      & RMSE   & CR (\%) \\
\midrule
\multicolumn{8}{l}{\textbf{Rate}} \\
$\theta_1$ & 5.14  & 5.1850 & 0.3586 & 0.8752 & 12.5435 & 0.3613 & 93.0 \\
$\theta_2$ & 10.67 & 10.6781& 0.5579 & 0.0760 & 1.4543  & 0.5577 & 95.6 \\
$\theta_3$ & 30.38 & 30.5466& 1.5370 & 0.5484 & 10.8393 & 1.5453 & 94.6 \\
$\theta_4$ & 50.02 & 50.0542& 2.4999 & 0.0684 & 1.3682  & 2.4989 & 94.8 \\
$\theta_5$ & 2.00  & 2.0167 & 0.1568 & 0.8350 & 10.6498 & 0.1576 & 92.6 \\
\midrule
\multicolumn{8}{l}{\textbf{Dispersion}} \\
$\lambda_1$ & 0.6445  & 0.6410 & 0.0264 & 0.5470 & 13.3458 & 0.0266 & 89.9 \\
$\lambda_2$ & 0.1420  & 0.1412 & 0.0423 & 0.5453 & 1.8320  & 0.0423 & 94.8 \\
$\lambda_3$ & -0.1378 & -0.1437& 0.0548 & 4.2765 & 10.7598 & 0.0551 & 95.1 \\
$\lambda_4$ & -0.0499 & -0.0505& 0.0513 & 1.1458 & 1.1152  & 0.0512 & 95.1 \\
$\lambda_5$ & 0.3650  & 0.3587 & 0.0441 & 1.7225 & 14.2519 & 0.0445 & 89.6 \\
\midrule
\multicolumn{8}{l}{\textbf{Correlation}} \\
$\rho_{12}$ & 0.0644  & 0.0644 & 0.0714 & 0.7764 & 0.7003 & 0.0713 & 95.0 \\
$\rho_{13}$ & 0.1041  & 0.1024 & 0.0710 & 1.6063 & 2.3541 & 0.0710 & 94.7 \\
$\rho_{14}$ & -0.0658 & -0.0642& 0.0684 & 2.5029 & 2.4063 & 0.0684 & 95.8 \\
$\rho_{15}$ & 0.2619  & 0.2646 & 0.0655 & 1.0197 & 4.0762 & 0.0655 & 94.9 \\
$\rho_{23}$ & 0.1008  & 0.0956 & 0.0708 & 5.1727 & 7.3601 & 0.0710 & 94.7 \\
$\rho_{24}$ & 0.1246  & 0.1285 & 0.0694 & 3.1008 & 5.5670 & 0.0695 & 96.1 \\
$\rho_{25}$ & -0.0122 & -0.0113& 0.0709 & 7.5507 & 1.3000 & 0.0708 & 95.3 \\
$\rho_{34}$ & 0.0867  & 0.0858 & 0.0693 & 1.0258 & 1.2825 & 0.0693 & 95.7 \\
$\rho_{35}$ & 0.1724  & 0.1728 & 0.0665 & 0.2265 & 0.5872 & 0.0665 & 94.7 \\
$\rho_{45}$ & 0.0452  & 0.0452 & 0.0710 & 0.0128 & 0.0082 & 0.0709 & 95.3 \\
\bottomrule
\end{tabular}
\end{table}

\begin{table}[H]
\centering
\caption{Results from the generalized Poisson distribution simulation in a large sample case ($N = 2000$).}
\label{tab:gp_large}
\scriptsize
\setlength{\tabcolsep}{5pt}
\renewcommand{\arraystretch}{0.92}

\begin{tabular}{lrrrrrrr}
\toprule
      & TV    & AE     & SD     & RB     & SB     & RMSE   & CR (\%) \\
\midrule
\multicolumn{8}{l}{\textbf{Rate}} \\
$\theta_1$ & 5.14  & 5.1462 & 0.1138 & 0.1201 & 5.4267 & 0.1139 & 95.4 \\
$\theta_2$ & 10.67 & 10.6656& 0.1836 & 0.0414 & 2.4079 & 0.1836 & 95.5 \\
$\theta_3$ & 30.38 & 30.3925& 0.4891 & 0.0412 & 2.5598 & 0.4890 & 95.1 \\
$\theta_4$ & 50.02 & 49.9694& 0.8170 & 0.1012 & 6.1970 & 0.8181 & 95.3 \\
$\theta_5$ & 2.00  & 1.9999 & 0.0504 & 0.0045 & 0.1779 & 0.0503 & 94.2 \\
\midrule
\multicolumn{8}{l}{\textbf{Dispersion}} \\
$\lambda_1$ & 0.6445  & 0.6442 & 0.0085 & 0.0429 & 3.2345 & 0.0085 & 94.4 \\
$\lambda_2$ & 0.1420  & 0.1423 & 0.0139 & 0.2143 & 2.1941 & 0.0139 & 95.4 \\
$\lambda_3$ & -0.1378 & -0.1381& 0.0179 & 0.2521 & 1.9374 & 0.0179 & 94.8 \\
$\lambda_4$ & -0.0499 & -0.0489& 0.0166 & 2.0454 & 6.1413 & 0.0166 & 95.2 \\
$\lambda_5$ & 0.3650  & 0.3654 & 0.0143 & 0.1129 & 2.8749 & 0.0143 & 94.0 \\
\midrule
\multicolumn{8}{l}{\textbf{Correlation}} \\
$\rho_{12}$ & 0.0644  & 0.0637 & 0.0221 & 1.0214 & 2.9766 & 0.0221 & 95.3 \\
$\rho_{13}$ & 0.1041  & 0.1038 & 0.0219 & 0.2403 & 1.1425 & 0.0219 & 96.2 \\
$\rho_{14}$ & -0.0658 & -0.0655& 0.0223 & 0.5072 & 1.4956 & 0.0223 & 95.4 \\
$\rho_{15}$ & 0.2619  & 0.2609 & 0.0198 & 0.3851 & 5.1067 & 0.0198 & 96.3 \\
$\rho_{23}$ & 0.1008  & 0.1004 & 0.0220 & 0.4172 & 1.9095 & 0.0220 & 94.8 \\
$\rho_{24}$ & 0.1246  & 0.1255 & 0.0223 & 0.6824 & 3.8124 & 0.0223 & 95.4 \\
$\rho_{25}$ & -0.0122 & -0.0128& 0.0219 & 5.0595 & 2.8194 & 0.0219 & 95.4 \\
$\rho_{34}$ & 0.0867  & 0.0854 & 0.0225 & 1.4723 & 5.6794 & 0.0225 & 94.8 \\
$\rho_{35}$ & 0.1724  & 0.1723 & 0.0212 & 0.0475 & 0.3854 & 0.0212 & 95.5 \\
$\rho_{45}$ & 0.0452  & 0.0456 & 0.0223 & 0.9839 & 1.9903 & 0.0223 & 96.2 \\
\bottomrule
\end{tabular}
\end{table}

\subsection{Multivariate Negative Binomial Distribution}

Next, we consider five negative binomial variables with numbers of successes
$\mathbf{r} = (3,\,8,\,15,\,20,\,43)$ and probabilities
$\mathbf{p} = (0.33,\,0.45,\,0.24,\,0.61,\,0.58)$.
Further, we specify the target correlation matrix $\Sigma$ using an exchangeable structure given below:
\[
\Sigma =
\begin{pmatrix}
1    & 0.50 & 0.50 & 0.50 & 0.50 \\
0.50 & 1    & 0.50 & 0.50 & 0.50 \\
0.50 & 0.50 & 1    & 0.50 & 0.50 \\
0.50 & 0.50 & 0.50 & 1    & 0.50 \\
0.50 & 0.50 & 0.50 & 0.50 & 1
\end{pmatrix}
\]

\begin{table}[H]
\centering
\caption{Results from the negative binomial distribution simulation in a small sample case ($N = 200$).}
\label{tab:nb_small}
\scriptsize
\setlength{\tabcolsep}{5pt}
\renewcommand{\arraystretch}{0.92}

\begin{tabular}{lrrrrrrr}
\toprule
      & TV    & AE     & SD     & RB     & SB      & RMSE   & CR (\%) \\
\midrule
\multicolumn{8}{l}{\textbf{Number of successes}} \\
$r_1$ & 3  & 3.0920  & 0.5322 & 3.0660 & 17.2838 & 0.5398 & 92.6 \\
$r_2$ & 8  & 8.3811  & 1.7083 & 4.7642 & 22.3115 & 1.7494 & 94.3 \\
$r_3$ & 15 & 15.3563 & 2.2285 & 2.3755 & 15.9899 & 2.2557 & 93.6 \\
$r_4$ & 20 & 21.3956 & 6.7713 & 6.9780 & 20.6108 & 6.9103 & 95.6 \\
$r_5$ & 43 & 45.8188 & 12.9911& 6.5553 & 21.6977 & 13.2870& 94.3 \\
\midrule
\multicolumn{8}{l}{\textbf{Probability}} \\
$p_1$ & 0.33 & 0.3347 & 0.0389 & 1.4201 & 12.0582 & 0.0391 & 91.8 \\
$p_2$ & 0.45 & 0.4577 & 0.0489 & 1.7213 & 15.8309 & 0.0495 & 93.9 \\
$p_3$ & 0.24 & 0.2434 & 0.0265 & 1.4271 & 12.9291 & 0.0267 & 92.5 \\
$p_4$ & 0.61 & 0.6144 & 0.0625 & 0.7218 & 7.0434  & 0.0626 & 94.2 \\
$p_5$ & 0.58 & 0.5857 & 0.0603 & 0.9911 & 9.5351  & 0.0605 & 93.9 \\
\midrule
\multicolumn{8}{l}{\textbf{Correlation}} \\
$\rho_{12}$ & 0.50 & 0.5026 & 0.0529 & 0.5105 & 4.8270 & 0.0529 & 94.7 \\
$\rho_{13}$ & 0.50 & 0.5039 & 0.0512 & 0.7749 & 7.5676 & 0.0513 & 95.3 \\
$\rho_{14}$ & 0.50 & 0.4981 & 0.0492 & 0.3866 & 3.9321 & 0.0492 & 96.4 \\
$\rho_{15}$ & 0.50 & 0.5028 & 0.0504 & 0.5551 & 5.5084 & 0.0504 & 96.1 \\
$\rho_{23}$ & 0.50 & 0.5005 & 0.0500 & 0.1033 & 1.0329 & 0.0500 & 96.3 \\
$\rho_{24}$ & 0.50 & 0.5003 & 0.0515 & 0.0659 & 0.6396 & 0.0515 & 95.5 \\
$\rho_{25}$ & 0.50 & 0.5025 & 0.0513 & 0.4979 & 4.8492 & 0.0514 & 95.2 \\
$\rho_{34}$ & 0.50 & 0.4993 & 0.0502 & 0.1482 & 1.4761 & 0.0502 & 95.2 \\
$\rho_{35}$ & 0.50 & 0.4982 & 0.0507 & 0.3644 & 3.5954 & 0.0507 & 96.2 \\
$\rho_{45}$ & 0.50 & 0.4993 & 0.0528 & 0.1418 & 1.3415 & 0.0528 & 94.7 \\
\bottomrule
\end{tabular}
\end{table}

\begin{table}[H]
\centering
\caption{ Results from the negative binomial distribution simulation in a large sample case ($N = 2000$).}
\label{tab:nb_large}
\scriptsize
\setlength{\tabcolsep}{5pt}
\renewcommand{\arraystretch}{0.92}

\begin{tabular}{lrrrrrrr}
\toprule
      & TV    & AE     & SD     & RB     & SB     & RMSE   & CR (\%) \\
\midrule
\multicolumn{8}{l}{\textbf{Number of successes}} \\
$r_1$ & 3  & 3.0116 & 0.1663 & 0.3860 & 6.9649 & 0.1666 & 95.0 \\
$r_2$ & 8  & 8.0225 & 0.5036 & 0.2813 & 4.4688 & 0.5039 & 94.7 \\
$r_3$ & 15 & 15.0072& 0.6380 & 0.0483 & 1.1344 & 0.6378 & 95.2 \\
$r_4$ & 20 & 20.0487& 1.6287 & 0.2435 & 2.9904 & 1.6287 & 95.0 \\
$r_5$ & 43 & 43.2188& 3.3430 & 0.5089 & 6.5457 & 3.3485 & 95.0 \\
\midrule
\multicolumn{8}{l}{\textbf{Probability}} \\
$p_1$ & 0.33 & 0.3310 & 0.0127 & 0.3012 & 7.8495 & 0.0127 & 94.0 \\
$p_2$ & 0.45 & 0.4503 & 0.0158 & 0.0765 & 2.1867 & 0.0157 & 94.3 \\
$p_3$ & 0.24 & 0.2400 & 0.0079 & 0.0097 & 0.2956 & 0.0079 & 94.9 \\
$p_4$ & 0.61 & 0.6097 & 0.0192 & 0.0515 & 1.6343 & 0.0192 & 95.0 \\
$p_5$ & 0.58 & 0.5805 & 0.0186 & 0.0793 & 2.4695 & 0.0186 & 95.4 \\
\midrule
\multicolumn{8}{l}{\textbf{Correlation}} \\
$\rho_{12}$ & 0.50 & 0.5005 & 0.0166 & 0.0949 & 2.8512  & 0.0166 & 96.0 \\
$\rho_{13}$ & 0.50 & 0.5006 & 0.0163 & 0.1115 & 3.4218  & 0.0163 & 95.8 \\
$\rho_{14}$ & 0.50 & 0.5013 & 0.0158 & 0.2582 & 8.1710  & 0.0158 & 96.1 \\
$\rho_{15}$ & 0.50 & 0.5009 & 0.0166 & 0.1815 & 5.4711  & 0.0166 & 95.3 \\
$\rho_{23}$ & 0.50 & 0.4995 & 0.0165 & 0.1039 & 3.1492  & 0.0165 & 95.9 \\
$\rho_{24}$ & 0.50 & 0.5019 & 0.0161 & 0.3783 & 11.7268 & 0.0162 & 94.8 \\
$\rho_{25}$ & 0.50 & 0.4983 & 0.0160 & 0.3479 & 10.8474 & 0.0161 & 95.8 \\
$\rho_{34}$ & 0.50 & 0.4994 & 0.0159 & 0.1283 & 4.0334  & 0.0159 & 95.4 \\
$\rho_{35}$ & 0.50 & 0.4988 & 0.0156 & 0.2392 & 7.6740  & 0.0156 & 96.3 \\
$\rho_{45}$ & 0.50 & 0.4992 & 0.0157 & 0.1534 & 4.8727  & 0.0158 & 96.0 \\
\bottomrule
\end{tabular}
\end{table}

\subsection{Multivariate Binomial Distribution}
In our third simulation scenario, we consider five binomial variables with numbers of trials
$\mathbf{n} = (5,\,12,\,25,\,30,\,40)$ and probabilities
$\mathbf{p} = (0.68,\,0.36,\,0.45,\,0.51,\,0.57)$.
Further, we specify the target correlation matrix $\Sigma$ using a Toeplitz structure:
\vspace{-0.5\baselineskip}
\[
\Sigma =
\begin{pmatrix}
1    & 0.45 & 0.40 & 0.35 & 0.30 \\
0.45 & 1    & 0.45 & 0.40 & 0.35 \\
0.40 & 0.45 & 1    & 0.45 & 0.40 \\
0.35 & 0.40 & 0.45 & 1    & 0.45 \\
0.30 & 0.35 & 0.40 & 0.45 & 1
\end{pmatrix}
\]
\vspace{-1.3\baselineskip}

\begin{table}[H]
\centering
\caption{ Results from the binomial distribution simulation in a small sample case ($N = 200$).}
\label{tab:bin_small}
\scriptsize
\setlength{\tabcolsep}{5pt}
\renewcommand{\arraystretch}{0.92}

\begin{tabular}{lrrrrrrr}
\toprule
      & TV    & AE     & SD     & RB     & SB      & RMSE   & CR (\%) \\
\midrule
\multicolumn{8}{l}{\textbf{Number of trials}} \\
$n_1$ & 5  & 5.0138  & 0.2126 & 0.2767 & 6.5088  & 0.2129 & 93.6 \\
$n_2$ & 12 & 12.5530 & 2.4353 & 4.6086 & 22.7095 & 2.4961 & 93.9 \\
$n_3$ & 25 & 25.4080 & 3.4450 & 1.6320 & 11.8429 & 3.4674 & 94.2 \\
$n_4$ & 30 & 30.0742 & 2.9018 & 0.2474 & 2.5575  & 2.9013 & 92.8 \\
$n_5$ & 40 & 40.3039 & 3.2077 & 0.7598 & 9.4752  & 3.2205 & 92.3 \\
\midrule
\multicolumn{8}{l}{\textbf{Probability}} \\
$p_1$ & 0.68 & 0.6795 & 0.0322 & 0.0665 & 1.4057 & 0.0322 & 94.0 \\
$p_2$ & 0.36 & 0.3554 & 0.0617 & 1.2717 & 7.4186 & 0.0618 & 93.9 \\
$p_3$ & 0.45 & 0.4504 & 0.0560 & 0.0809 & 0.6504 & 0.0559 & 94.1 \\
$p_4$ & 0.51 & 0.5136 & 0.0477 & 0.6990 & 7.4744 & 0.0478 & 93.2 \\
$p_5$ & 0.57 & 0.5692 & 0.0438 & 0.1361 & 1.7698 & 0.0438 & 92.8 \\
\midrule
\multicolumn{8}{l}{\textbf{Correlation}} \\
$\rho_{12}$ & 0.45 & 0.4528 & 0.0554 & 0.6112 & 4.9628 & 0.0555 & 94.7 \\
$\rho_{13}$ & 0.40 & 0.4026 & 0.0573 & 0.6596 & 4.6024 & 0.0574 & 95.1 \\
$\rho_{14}$ & 0.35 & 0.3496 & 0.0609 & 0.1188 & 0.6824 & 0.0609 & 95.0 \\
$\rho_{15}$ & 0.30 & 0.2970 & 0.0676 & 1.0141 & 4.5025 & 0.0676 & 93.8 \\
$\rho_{23}$ & 0.45 & 0.4501 & 0.0547 & 0.0328 & 0.2697 & 0.0546 & 95.2 \\
$\rho_{24}$ & 0.40 & 0.4003 & 0.0573 & 0.0682 & 0.4761 & 0.0573 & 95.4 \\
$\rho_{25}$ & 0.35 & 0.3511 & 0.0629 & 0.3107 & 1.7293 & 0.0629 & 94.2 \\
$\rho_{34}$ & 0.45 & 0.4514 & 0.0533 & 0.3019 & 2.5506 & 0.0532 & 95.8 \\
$\rho_{35}$ & 0.40 & 0.4048 & 0.0578 & 1.2098 & 8.3653 & 0.0580 & 95.2 \\
$\rho_{45}$ & 0.45 & 0.4511 & 0.0552 & 0.2456 & 2.0026 & 0.0552 & 94.7 \\
\bottomrule
\end{tabular}
\end{table}

\begin{table}[H]
\centering
\caption{ Results from the binomial distribution simulation in the large sample case ($N = 2000$).}
\label{tab:bin_large}
\scriptsize
\setlength{\tabcolsep}{5pt}
\renewcommand{\arraystretch}{0.92}

\begin{tabular}{lrrrrrrr}
\toprule
      & TV    & AE     & SD     & RB     & SB     & RMSE   & CR (\%) \\
\midrule
\multicolumn{8}{l}{\textbf{Number of trials}} \\
$n_1$ & 5  & 5.0013  & 0.0656 & 0.0267 & 2.0306 & 0.0656 & 95.3 \\
$n_2$ & 12 & 12.0496 & 0.6334 & 0.4137 & 7.8373 & 0.6350 & 95.7 \\
$n_3$ & 25 & 25.0123 & 0.9441 & 0.0493 & 1.3055 & 0.9437 & 94.9 \\
$n_4$ & 30 & 29.9948 & 0.8770 & 0.0174 & 0.5949 & 0.8765 & 94.7 \\
$n_5$ & 40 & 40.0240 & 0.9753 & 0.0599 & 2.4586 & 0.9751 & 94.1 \\
\midrule
\multicolumn{8}{l}{\textbf{Probability}} \\
$p_1$ & 0.68 & 0.6798 & 0.0102 & 0.0255 & 1.6923 & 0.0102 & 95.7 \\
$p_2$ & 0.36 & 0.3596 & 0.0190 & 0.1001 & 1.8951 & 0.0190 & 95.4 \\
$p_3$ & 0.45 & 0.4504 & 0.0170 & 0.0861 & 2.2807 & 0.0170 & 94.9 \\
$p_4$ & 0.51 & 0.5106 & 0.0151 & 0.1212 & 4.0901 & 0.0151 & 94.9 \\
$p_5$ & 0.57 & 0.5710 & 0.0141 & 0.0057 & 0.2328 & 0.0141 & 94.8 \\
\midrule
\multicolumn{8}{l}{\textbf{Correlation}} \\
$\rho_{12}$ & 0.45 & 0.4497 & 0.0172 & 0.0692 & 1.8135  & 0.0172 & 96.7 \\
$\rho_{13}$ & 0.40 & 0.4013 & 0.0190 & 0.3282 & 6.9095  & 0.0190 & 95.2 \\
$\rho_{14}$ & 0.35 & 0.3520 & 0.0195 & 0.0024 & 0.0431  & 0.0195 & 95.3 \\
$\rho_{15}$ & 0.30 & 0.2993 & 0.0206 & 0.2214 & 3.2287  & 0.0206 & 94.9 \\
$\rho_{23}$ & 0.45 & 0.4481 & 0.0171 & 0.4224 & 11.1115 & 0.0172 & 95.7 \\
$\rho_{24}$ & 0.40 & 0.3985 & 0.0187 & 0.3729 & 7.9805  & 0.0187 & 95.4 \\
$\rho_{25}$ & 0.35 & 0.3485 & 0.0194 & 0.4196 & 7.5808  & 0.0194 & 95.6 \\
$\rho_{34}$ & 0.45 & 0.4486 & 0.0179 & 0.3057 & 7.6891  & 0.0179 & 95.4 \\
$\rho_{35}$ & 0.40 & 0.4004 & 0.0185 & 0.0900 & 1.9420  & 0.0185 & 95.1 \\
$\rho_{45}$ & 0.45 & 0.4497 & 0.0173 & 0.0686 & 1.7886  & 0.0172 & 94.8 \\
\bottomrule
\end{tabular}
\end{table}

\subsection{Mixed Generalized Poisson, Negative Binomial, and Binomial Distributions}

In the last simulation scenario, we consider a six-variate mixed distribution consisting of two
generalized Poisson variables with $\boldsymbol{\theta}_{\mathrm{GPD}} = (9.39,\,18.6)$ and
$\boldsymbol{\lambda}_{\mathrm{GPD}} = (-0.023,\,0.1203)$, two negative binomial variables
$\boldsymbol{r}_{\mathrm{NB}} = (6,\,15)$ and $\boldsymbol{p}_{\mathrm{NB}} = (0.54,\,0.47)$,
and two binomial variables with $\boldsymbol{n}_{\mathrm{B}} = (20,\,40)$ and
$\boldsymbol{p}_{\mathrm{B}} = (0.62,\,0.58)$. We then specify the target correlation matrix as follows:
\[
\Sigma =
\begin{pmatrix}
1    & 0.28 & 0.31 & 0.27 & 0.24 & 0.17 \\
0.28 & 1    & 0.18 & 0.26 & 0.11 & 0.12 \\
0.31 & 0.18 & 1    & 0.14 & 0.23 & 0.26 \\
0.27 & 0.26 & 0.14 & 1    & 0.24 & 0.13 \\
0.24 & 0.11 & 0.23 & 0.24 & 1    & 0.15 \\
0.17 & 0.12 & 0.26 & 0.13 & 0.15 & 1
\end{pmatrix}
\]

\begin{table}[H]
\centering
\caption{ Results from the mixed distribution simulation in the small sample case ($N = 200$).}
\label{tab:mixed_small}
\scriptsize
\setlength{\tabcolsep}{5pt}
\renewcommand{\arraystretch}{0.92}

\begin{tabular}{lrrrrrrr}
\toprule
      & TV    & AE     & SD     & RB     & SB      & RMSE   & CR (\%) \\
\midrule
\multicolumn{8}{l}{\textbf{ Parameters }} \\
$\theta_{\mathrm{GPD}1}$  & 9.39   & 9.4042  & 0.5244 & 0.1510 & 2.7032  & 0.5243 & 94.8 \\
$\theta_{\mathrm{GPD}2}$  & 18.6   & 18.6913 & 1.0127 & 0.4911 & 9.0196  & 1.0163 & 93.6 \\
$\lambda_{\mathrm{GPD}1}$ & -0.023 & -0.0239 & 0.0517 & 3.9775 & 1.7707  & 0.0516 & 94.0 \\
$\lambda_{\mathrm{GPD}2}$ & 0.1203 & 0.1156  & 0.0460 & 3.9040 & 10.2125 & 0.0462 & 92.4 \\
$r_{\mathrm{NB}1}$        & 6      & 6.4210  & 1.6009 & 7.0173 & 26.2995 & 1.6546 & 94.2 \\
$r_{\mathrm{NB}2}$        & 15     & 15.7855 & 3.3845 & 5.2367 & 23.2088 & 3.4728 & 93.0 \\
$p_{\mathrm{NB}1}$        & 0.54   & 0.5490  & 0.0579 & 1.6637 & 15.5048 & 0.0586 & 92.7 \\
$p_{\mathrm{NB}2}$        & 0.47   & 0.4779  & 0.0509 & 1.6787 & 15.4860 & 0.0515 & 92.4 \\
$n_{\mathrm{B}1}$         & 20     & 20.0846 & 1.2005 & 0.4231 & 7.0493  & 1.2029 & 95.0 \\
$n_{\mathrm{B}2}$         & 40     & 40.1777 & 2.9639 & 0.4444 & 5.9970  & 2.9677 & 93.5 \\
$p_{\mathrm{B}1}$         & 0.62   & 0.6199  & 0.0374 & 0.0222 & 0.3674  & 0.0374 & 94.9 \\
$p_{\mathrm{B}2}$         & 0.58   & 0.5805  & 0.0422 & 0.0844 & 1.1611  & 0.0421 & 93.6 \\
\midrule
\multicolumn{8}{l}{\textbf{Correlation}} \\
$\rho_{12}$ & 0.28 & 0.2810 & 0.0655 & 0.3576 & 1.5287 & 0.0655 & 95.3 \\
$\rho_{13}$ & 0.31 & 0.3107 & 0.0637 & 0.2330 & 1.1337 & 0.0637 & 94.0 \\
$\rho_{14}$ & 0.27 & 0.2710 & 0.0639 & 0.3840 & 1.6217 & 0.0639 & 95.4 \\
$\rho_{15}$ & 0.24 & 0.2388 & 0.0685 & 0.4991 & 1.7498 & 0.0684 & 94.2 \\
$\rho_{16}$ & 0.17 & 0.1724 & 0.0670 & 1.4051 & 3.5648 & 0.0670 & 94.8 \\
$\rho_{23}$ & 0.18 & 0.1818 & 0.0702 & 1.0037 & 2.5737 & 0.0702 & 93.8 \\
$\rho_{24}$ & 0.26 & 0.2609 & 0.0676 & 0.3529 & 1.3576 & 0.0676 & 94.3 \\
$\rho_{25}$ & 0.11 & 0.1106 & 0.0693 & 0.5178 & 0.8224 & 0.0692 & 94.8 \\
$\rho_{26}$ & 0.12 & 0.1163 & 0.0705 & 3.0521 & 5.1915 & 0.0706 & 95.0 \\
$\rho_{34}$ & 0.14 & 0.1435 & 0.0688 & 2.5166 & 5.1191 & 0.0689 & 94.8 \\
$\rho_{35}$ & 0.23 & 0.2312 & 0.0681 & 0.5175 & 1.7466 & 0.0681 & 95.0 \\
$\rho_{36}$ & 0.26 & 0.2545 & 0.0671 & 2.1081 & 8.1705 & 0.0673 & 94.7 \\
$\rho_{45}$ & 0.24 & 0.2412 & 0.0671 & 0.4914 & 1.7590 & 0.0670 & 95.3 \\
$\rho_{46}$ & 0.13 & 0.1333 & 0.0670 & 2.5193 & 4.8890 & 0.0670 & 95.6 \\
$\rho_{56}$ & 0.15 & 0.1512 & 0.0698 & 0.7834 & 1.6846 & 0.0697 & 93.8 \\
\bottomrule
\end{tabular}
\end{table}

\begin{table}[H]
\centering
\caption{ Results from the mixed distribution simulation in the large sample case ($N = 2000$).}
\label{tab:mixed_large}
\scriptsize
\setlength{\tabcolsep}{5pt}
\renewcommand{\arraystretch}{0.92}

\begin{tabular}{lrrrrrrr}
\toprule
      & TV    & AE     & SD     & RB     & SB      & RMSE   & CR (\%) \\
\midrule
\multicolumn{8}{l}{\textbf{ Parameters }} \\
$\theta_{\mathrm{GPD}1}$  & 9.39   & 9.3957  & 0.1643 & 0.0605 & 3.4588  & 0.1643 & 94.9 \\
$\theta_{\mathrm{GPD}2}$  & 18.6   & 18.6069 & 0.2915 & 0.0369 & 2.3531  & 0.2915 & 96.4 \\
$\lambda_{\mathrm{GPD}1}$ & -0.023 & -0.0232 & 0.0161 & 0.8529 & 1.2185  & 0.0161 & 94.3 \\
$\lambda_{\mathrm{GPD}2}$ & 0.1203 & 0.1202  & 0.0132 & 0.0789 & 0.7181  & 0.0132 & 96.0 \\
$r_{\mathrm{NB}1}$        & 6      & 6.0582  & 0.4708 & 0.9708 & 12.3714 & 0.4742 & 93.2 \\
$r_{\mathrm{NB}2}$        & 15     & 15.0741 & 0.9823 & 0.4941 & 7.5454  & 0.9846 & 93.4 \\
$p_{\mathrm{NB}1}$        & 0.54   & 0.5416  & 0.0194 & 0.2911 & 8.1072  & 0.0194 & 92.6 \\
$p_{\mathrm{NB}2}$        & 0.47   & 0.4707  & 0.0162 & 0.1451 & 4.1960  & 0.0163 & 93.6 \\
$n_{\mathrm{B}1}$         & 20     & 20.0167 & 0.3820 & 0.0834 & 4.3648  & 0.3822 & 94.8 \\
$n_{\mathrm{B}2}$         & 40     & 40.0517 & 0.8857 & 0.1294 & 5.8420  & 0.8867 & 95.3 \\
$p_{\mathrm{B}1}$         & 0.62   & 0.6197  & 0.0119 & 0.0437 & 2.2677  & 0.0119 & 94.7 \\
$p_{\mathrm{B}2}$         & 0.58   & 0.5796  & 0.0129 & 0.0604 & 2.7217  & 0.0129 & 95.3 \\
\midrule
\multicolumn{8}{l}{\textbf{Correlation}} \\
$\rho_{12}$ & 0.28 & 0.2785 & 0.0195 & 0.5228 & 7.5088 & 0.0195 & 96.5 \\
$\rho_{13}$ & 0.31 & 0.3089 & 0.0198 & 0.3415 & 5.3443 & 0.0198 & 96.2 \\
$\rho_{14}$ & 0.27 & 0.2689 & 0.0203 & 0.3944 & 5.2374 & 0.0204 & 95.4 \\
$\rho_{15}$ & 0.24 & 0.2394 & 0.0214 & 0.2380 & 2.6682 & 0.0214 & 93.2 \\
$\rho_{16}$ & 0.17 & 0.1689 & 0.0224 & 0.6206 & 4.7089 & 0.0224 & 94.2 \\
$\rho_{23}$ & 0.18 & 0.1789 & 0.0218 & 0.5856 & 4.8447 & 0.0218 & 94.9 \\
$\rho_{24}$ & 0.26 & 0.2613 & 0.0212 & 0.5074 & 6.2319 & 0.0212 & 94.4 \\
$\rho_{25}$ & 0.11 & 0.1101 & 0.0214 & 0.0924 & 0.4745 & 0.0214 & 95.8 \\
$\rho_{26}$ & 0.12 & 0.1186 & 0.0224 & 1.1851 & 6.3370 & 0.0225 & 94.6 \\
$\rho_{34}$ & 0.14 & 0.1397 & 0.0228 & 0.2221 & 1.3612 & 0.0228 & 93.7 \\
$\rho_{35}$ & 0.23 & 0.2283 & 0.0213 & 0.7314 & 7.8995 & 0.0214 & 95.0 \\
$\rho_{36}$ & 0.26 & 0.2590 & 0.0212 & 0.3917 & 4.7985 & 0.0212 & 94.6 \\
$\rho_{45}$ & 0.24 & 0.2406 & 0.0204 & 0.2482 & 2.9165 & 0.0204 & 95.7 \\
$\rho_{46}$ & 0.13 & 0.1303 & 0.0224 & 0.1999 & 1.1581 & 0.0224 & 94.9 \\
$\rho_{56}$ & 0.15 & 0.1508 & 0.0222 & 0.5495 & 3.7184 & 0.0222 & 95.3 \\
\bottomrule
\end{tabular}
\end{table}

Tables 1--8 summarize the results for the four data generation scenarios. In the small-sample
scenarios ($N = 200$), the discrepancies between the average estimates (AE) and the true values
(TV) are minimal across all parameters of interest, particularly when the parameter values are
large. For the large sample size scenario ($N = 2000$), the average estimates are consistently close
to their corresponding true values. In both small and large sample size scenarios, the relative bias
(RB) values are generally below the 5\% acceptable threshold. Standardized bias (SB) and RMSE
values naturally decrease as sample size increases, demonstrating improved precision and reduced
sampling variability under larger samples. Moreover, the coverage rates (CR) remain around the
nominal 95\% level across all scenarios. Overall, these results provide strong evidence that our
proposed method can provide accurate and stable parameter estimation across a wide range of
settings.

\section{R Package: MultiDiscreteRNG}
This package implements a general framework of the algorithm in Section~2 to generate data that
follow multivariate generalized Poisson, negative binomial, binomial distributions, and mixed
discrete distributions. The package is available at
\url{https://github.com/ckchengtommy/MultiDiscreteRNG}.

The validation functions ensure that all user inputs are defined correctly before simulation.
\textit{validation.Bparameters}, \textit{validation.NBparameters}, and \textit{validation.GPDparameters}
check the feasibility of input parameters for the generalized Poisson, negative binomial, and
binomial distributions. For example, they verify parameter ranges (e.g., trial counts are
nonnegative integers; probabilities fall within 0 to 1; generalized Poisson parameter constraints
such as $\theta>0$ and $\lambda<1$).

Collapsing functions compute binary probabilities and thresholds from the specified discrete
margins. The functions \textit{calc.bin.prob.B}, \textit{calc.bin.prob.NB}, and
\textit{calc.bin.prob.GPD} convert each discrete margin to a binary variable. For each corresponding
distribution, the function computes its probability mass function over support values, calculates a
median threshold at which the distribution is dichotomized. It then returns the binary success
probability $p$, the probability mass function objects, and the threshold locations. These outputs
are later used in the reverse-collapsing step to reconstruct the original discrete scales.

The function \textit{discrete\_cont} computes the tetrachoric correlation matrix for a multivariate
standard normal distribution as part of Step~2 in the algorithm. If the resulting correlation matrix
is not positive definite, the function returns the nearest positive definite matrix and issues a
warning.

The \textit{generate.binaryVar} function produces multivariate binary data based on specified marginal
probabilities and a correlation matrix. It generates multivariate normal data using a tetrachoric
correlation matrix, which is calculated by the \textit{discrete\_cont} function, followed by
dichotomization according to the vector of binary probabilities.

The \textit{BinToMix} function maps the correlated binary variables back to their original discrete
families using the previously obtained probability mass functions and dichotomization locations.
They perform the reverse-collapsing step when the marginals are mixed. Each function returns the
simulated multivariate data together with its empirical correlation matrix.

The function \textit{simBinaryCorr.Mix} iteratively computes an intermediate binary correlation matrix
to achieve the specified correlation structure for discrete data. These functions repeatedly call
\textit{generate.binaryVar} and \textit{BinToMix} to align the correlation of the generated binary data
with the target. The \textit{simBinaryCorr.Mix} function extends these capabilities to generate mixed
discrete data with varied marginal distributions.

The \textit{genMix} function serves as the engine function for multivariate discrete data from binary
parameters. The output is a list consisting of two components: a matrix of multivariate discrete
data and its corresponding correlation matrix. Below provides a demonstration of the code in R.

\begin{lstlisting}[language=R]
library(MultiDiscreteRNG)
set.seed(2345)
# Specify Generalized Poisson, Negative binomial and binomial parameters
GPD.theta.vec = c(15.86709 , 7.50268)
GPD.lambda.vec = c(0.01862897, 0.0935569)
NB.r.vec = c(15, 20)
NB.prob.vec = c(0.45, 0.53)
B.n.vec = c(8, 16)
B.prob.vec = c(0.61, 0.72)

# Specify correlation matrix
M <- c(0.223, 0.135, 0.213, 0.203, 0.162, 0.114, 0.142,
       0.112, 0.131, 0.105, 0.156, 0.211, 0.158, 0.132, 0.183)
N <- diag(6)
N[lower.tri(N)] <- M
cmat <- N + t(N)
diag(cmat) <- 1

# Validate parameters
validation.GPDparameters(GPD.theta.vec, GPD.lambda.vec)
validation.NBparameters(NB.r.vec, NB.prob.vec)
validation.Bparameters(B.n.vec, B.prob.vec)

# Compute binary correlation
BinObjMix = simBinaryCorr.Mix(
  GPD.theta.vec = GPD.theta.vec,
  GPD.lambda.vec = GPD.lambda.vec,
  NB.r.vec = NB.r.vec,
  NB.prob.vec = NB.prob.vec,
  B.n.vec = B.n.vec,
  B.prob.vec = B.prob.vec,
  CorrMat = cmat,
  no.rows = 1000000
)

# Generate mixed discrete data with 2000 observations
MixData = genMix(no.rows = 2000, binObj = BinObjMix)$y

# Check the correlations of the generated data
round(cor(MixData), 5)
#          [,1]    [,2]    [,3]    [,4]    [,5]    [,6]
# [1,] 1.00000 0.22521 0.11737 0.23959 0.18407 0.16043
# [2,] 0.22521 1.00000 0.06646 0.13755 0.08582 0.10509
# [3,] 0.11737 0.06646 1.00000 0.10929 0.15800 0.20128
# [4,] 0.23959 0.13755 0.10929 1.00000 0.16663 0.14089
# [5,] 0.18407 0.08582 0.15800 0.16663 1.00000 0.15964
# [6,] 0.16043 0.10509 0.20128 0.14089 0.15964 1.00000
\end{lstlisting}

\subsection{Computational issues and implementation details }

When implementing the algorithm, several numerical issues may occur. First, the specified
pairwise correlation must conform to the correlation bounds, which can be determined using
the GSC sorting algorithm by Demirtas and Hedeker (2011). Any violations should be identified
and resolved before proceeding with the computation. Second, the calculated intermediate
correlation matrix in Step 3 does not guarantee positive definiteness. If this happens, the
nearest positive definite matrix can be computed using the methods suggested by Higham (2002),
implemented in the \textit{nearPD} function in the \textit{Matrix} package in R.

In addition, our approach relies on the intermediate step in generating multivariate binary
variables. For binary variable pairs, Emrich and Piedmonte (1991) demonstrated that the
correlation boundaries are imposed by the marginal expectations, where $\delta_{ij}^b$ is
bounded by
\[
\left\{
\max\!\left(
-\sqrt{\frac{p_i p_j}{q_i q_j}},\;
-\sqrt{\frac{q_i q_j}{p_i p_j}}
\right),\;
\min\!\left(
\sqrt{\frac{p_i q_j}{q_i p_j}},\;
\sqrt{\frac{q_i p_j}{p_i q_j}}
\right)
\right\},
\]
for $q_i = 1 - p_i$ and $q_j = 1 - p_j$. If the binary correlations go beyond these limits,
the results generated would be unreasonable. One example would be a specification of a
trivariate generalized Poisson distribution with
$\boldsymbol{\theta} = (23, 40, 4.6)$, $\boldsymbol{\lambda} = (0.72, 0.58, 0.14)$ and
correlations $\rho_{12}=0.6$, $\rho_{13}=0.24$, $\rho_{23}=0.71$. This yields a marginal
expectation of the intermediate binary variables $p_1^b=0.4545$, $p_2^b=0.7168$,
$p_3^b=0.5017$, and the corresponding upper bound $\delta_{12}^b=0.5738$,
$\delta_{13}^b=0.9099$, $\delta_{23}^b=0.6306$. Clearly, the specified correlations
$\rho_{12}=0.6$ and $\rho_{23}=0.71$ exceed these limits. Thus, the proposed method becomes
inapplicable in this scenario.

\section{Real Data Illustration }
Over-dispersion and under-dispersion are common issues in many real-world datasets. To address these features, we illustrate our proposed method for generating multivariate generalized Poisson, negative binomial data, and mixed discrete data. Specifically, we present three scenarios: the generation of multivariate generalized Poisson data, multivariate negative binomial data, and multivariate mixed data with components of both generalized Poisson and negative binomial distributions.

\subsection{National Medical Expenditure Survey (NMES)}

The dataset originated from the National Medical Expenditure Survey (NMES), conducted in
1987 and 1988, providing a detailed look at the usage and funding of health services by
Americans covered by Medicare. Deb and Trivedi (1997) analyzed this dataset, focusing on
4,406 individuals aged 66 and older who benefit from Medicare (a public insurance program).
The data are available in the \textit{DebTrivedi.rda} file from the R package \textit{MixAll}
by Iovleff (2019).

We illustrate the multivariate generalized Poisson data generation method using five count
variables: visits to a physician in an office setting (OFP), number of emergency room visits
(EMER), visits to a physician in a hospital outpatient setting (OPP) adjusted by adding 1 to
avoid computational complexities (Li et al., 2021), the number of chronic diseases and
conditions (NUMCHRON), and the years of education received (SCHOOL). To closely mimic our
simulation setting, we select participants who are 80 years old or older from the dataset to
yield a sample size $N_{\text{sample}} = 190$. To evaluate the performance of our data
generation mechanism, we created simulated data with a sample size of $N = 200$ over 1000
replications to recover the parameters and pairwise correlations of the NMES dataset. We
apply the method of moments to the dataset to obtain the true values of the rate $\theta$ and
dispersion $\lambda$ parameters of the marginal distribution. For the specified correlation
matrix, we consider the empirical Pearson correlation of the selected variables. The results
of the evaluation are presented in Table~9. The results show that the average estimates (AE)
align closely with the true values (TV), and the relative bias (RB) values remain within
acceptable limits, indicating good point estimation performance. While certain standardized
bias (SB) values are moderately elevated and several coverage rates (CR) fall marginally below
the nominal 95\% level, the overall precision and accuracy of parameter estimation remain
acceptable. These results demonstrate adequate performance in real-data applications, even
under conditions of small sample size.

\begin{table}[H]
\centering
\caption{ Results from the NMES dataset replication study ($N = 200$).}
\label{tab:nmes_replication}
\scriptsize
\setlength{\tabcolsep}{5pt}
\renewcommand{\arraystretch}{0.92}

\begin{tabular}{lrrrrrrr}
\toprule
      & TV    & AE     & SD     & RB     & SB     & RMSE   & CR (\%) \\
\midrule
\multicolumn{8}{l}{\textbf{Rate}} \\
$\theta_{1(\mathrm{OFP})}$      & 1.8604 & 1.8900 & 0.1630 & 1.5907 & 18.1522 & 0.1656 & 94.0 \\
$\theta_{2(\mathrm{EMER})}$     & 0.3322 & 0.3356 & 0.0442 & 1.0327 & 7.7632  & 0.0443 & 94.9 \\
$\theta_{3(\mathrm{OPP}+1)}$    & 1.2549 & 1.2656 & 0.1006 & 0.8536 & 10.6461 & 0.1011 & 95.3 \\
$\theta_{4(\mathrm{NUMCHRON})}$ & 1.4579 & 1.4614 & 0.1078 & 0.2390 & 3.2311  & 0.1078 & 96.1 \\
$\theta_{5(\mathrm{SCHOOL})}$   & 5.5628 & 5.5901 & 0.3298 & 0.4904 & 8.2720  & 0.3308 & 94.8 \\
\midrule
\multicolumn{8}{l}{\textbf{Dispersion}} \\
$\lambda_{1(\mathrm{OFP})}$      & 0.6158 & 0.6080 & 0.0377 & 1.2642 & 20.6633 & 0.0385 & 85.6 \\
$\lambda_{2(\mathrm{EMER})}$     & 0.1695 & 0.1581 & 0.0642 & 6.7315 & 17.7716 & 0.0652 & 84.8 \\
$\lambda_{3(\mathrm{OPP}+1)}$    & 0.0900 & 0.0846 & 0.0492 & 5.9493 & 10.8863 & 0.0494 & 91.6 \\
$\lambda_{4(\mathrm{NUMCHRON})}$ & 0.1064 & 0.1050 & 0.0463 & 1.3183 & 3.0306  & 0.0463 & 94.0 \\
$\lambda_{5(\mathrm{SCHOOL})}$   & 0.3606 & 0.3581 & 0.0364 & 0.6829 & 6.7691  & 0.0364 & 93.2 \\
\midrule
\multicolumn{8}{l}{\textbf{Correlation}} \\
$\rho_{12}$ & 0.2050 & 0.2122 & 0.0747 & 3.4846 & 9.5583 & 0.0750 & 91.7 \\
$\rho_{13}$ & 0.0303 & 0.0288 & 0.0687 & 5.2022 & 2.2964 & 0.0687 & 94.9 \\
$\rho_{14}$ & 0.0926 & 0.0928 & 0.0735 & 0.1720 & 0.2166 & 0.0735 & 93.2 \\
$\rho_{15}$ & 0.2320 & 0.2344 & 0.0723 & 1.0530 & 3.3780 & 0.0723 & 92.7 \\
$\rho_{23}$ & 0.2770 & 0.2800 & 0.0708 & 1.0511 & 4.1128 & 0.0708 & 92.2 \\
$\rho_{24}$ & 0.1191 & 0.1193 & 0.0718 & 0.1782 & 0.2956 & 0.0718 & 94.6 \\
$\rho_{25}$ & 0.0102 & 0.0106 & 0.0697 & 4.0640 & 0.5955 & 0.0697 & 95.0 \\
$\rho_{34}$ & 0.0253 & 0.0277 & 0.0684 & 9.4862 & 3.5573 & 0.0684 & 95.2 \\
$\rho_{35}$ & 0.0734 & 0.0715 & 0.0711 & 2.5893 & 2.6754 & 0.0710 & 94.3 \\
$\rho_{45}$ & 0.0698 & 0.0685 & 0.0692 & 1.9418 & 1.9578 & 0.0692 & 95.5 \\
\bottomrule
\end{tabular}
\end{table}

\subsection{Australian Health Survey }
The Australian Health Survey (AHS) represents the largest health-related survey conducted in Australia. The dataset is available in the \textit{ahs.Rdata} file from the R package \textit{mcglm} by Bonat (2018). The primary goal of this survey was to investigate whether increased access to healthcare services and demographic factors, such as sex, age, and income, are associated with the frequency of healthcare service utilization. The dataset includes 5,190 respondents from a cross-sectional study of individuals aged 18 years or older. We illustrate the multivariate negative binomial data generation method of four variables: number of consultations with a doctor or specialist (\texttt{Ndoc}), number of illnesses in the past 2 weeks (\texttt{ill}), number of admissions to a hospital, psychiatric hospital, nursing or convalescence home in the past 12 months (\texttt{Nadm}), and total number of prescribed and non-prescribed medications used in the past two days (\texttt{Nmed}). Adhering closely to our simulation setting, we selected participants under the age of 22 with complete data, resulting in a sample size of $N_{\textit{sample}} = 1965$. For each marginal distribution, we apply the method of moments to the dataset to obtain the true values of the number of successes $r$ (rounded up to an integer) and probability $p$ parameters. We obtain the specified correlation matrix using the empirical Pearson correlation of the selected variables. We generate simulated data with a sample of $N = 2000$ through 1000 replications to recover the target marginal distribution parameters and pairwise correlation values of the AHS dataset. The results are presented in Table~10. The results show that our method recovers the parameter of interest with reasonable accuracy, as reflected by the small differences between the true values (TV) and average estimates (AE). Despite some standardized bias (SB) values being higher, the relative bias values (RB) fall within acceptable limits, and most coverage rates (CR) are reasonably close to the nominal 95\% level.

\begin{table}[H]
\centering
\caption{ Results from the AHS dataset replication study ($N = 2000$).}
\label{tab:ahs_replication}
\scriptsize
\setlength{\tabcolsep}{5pt}
\renewcommand{\arraystretch}{0.92}

\begin{tabular}{lrrrrrrr}
\toprule
      & TV     & AE     & SD     & RB     & SB      & RMSE   & CR (\%) \\
\midrule
\multicolumn{8}{l}{\textbf{Number of successes}} \\
$r_{1(\mathrm{Ndoc})}$ & 1 & 1.0105 & 0.1002 & 1.0531  & 10.5064 & 0.1007 & 93.4 \\
$r_{2(\mathrm{ill})}$  & 7 & 7.3774 & 1.6893 & 5.3915  & 22.3414 & 1.7301 & 94.3 \\
$r_{3(\mathrm{Nadm})}$ & 1 & 1.1120 & 0.3988 & 11.1979 & 28.0785 & 0.4140 & 94.8 \\
$r_{4(\mathrm{Nmed})}$ & 2 & 2.0330 & 0.2739 & 1.6518  & 12.0612 & 0.2757 & 93.5 \\
\midrule
\multicolumn{8}{l}{\textbf{Probability}} \\
$p_{1(\mathrm{Ndoc})}$ & 0.5465 & 0.5475 & 0.0254 & 0.1985 & 4.2711 & 0.0254 & 94.2 \\
$p_{2(\mathrm{ill})}$  & 0.8349 & 0.8368 & 0.0281 & 0.2214 & 6.5842 & 0.0281 & 94.3 \\
$p_{3(\mathrm{Nadm})}$ & 0.8467 & 0.8491 & 0.0381 & 0.2861 & 6.3621 & 0.0381 & 92.7 \\
$p_{4(\mathrm{Nmed})}$ & 0.6969 & 0.6974 & 0.0280 & 0.0744 & 1.8477 & 0.0280 & 94.0 \\
\midrule
\multicolumn{8}{l}{\textbf{Correlation}} \\
$\rho_{12}$ & 0.1552 & 0.1556 & 0.0227 & 0.2784 & 1.8996 & 0.0227 & 93.5 \\
$\rho_{13}$ & 0.1085 & 0.1070 & 0.0230 & 1.4287 & 6.7331 & 0.0231 & 93.5 \\
$\rho_{14}$ & 0.1060 & 0.1055 & 0.0235 & 0.4301 & 1.9393 & 0.0235 & 94.2 \\
$\rho_{23}$ & 0.1952 & 0.1948 & 0.0219 & 0.2076 & 1.8522 & 0.0219 & 94.4 \\
$\rho_{24}$ & 0.2806 & 0.2793 & 0.0216 & 0.4383 & 5.6925 & 0.0216 & 93.5 \\
$\rho_{34}$ & 0.0520 & 0.0523 & 0.0235 & 0.5311 & 1.1742 & 0.0235 & 93.6 \\
\bottomrule
\end{tabular}
\end{table}

\subsection{Los Angeles Crime Data }
The Los Angeles (LA) Crime dataset, compiled from official reports and publicly available on Kaggle
(\url{https://www.kaggle.com/datasets/venkatsairo4899/los-angeles-crime-data-2020-2023}),
provides a comprehensive record of crime incidents in Los Angeles. For this study, we analyze
offense frequencies across areas and time periods. We aggregate records into an area name by
year--month panel that counts incidents for five categories of criminal offenses: grand theft (GT),
petty theft (PT), extortion (E), violation of court order (VCO), and intimate partner simple assault (IPSA),
yielding a final dataset with $N_{\mathrm{sample}} = 1197$ observations. To illustrate the generation of
multivariate mixed discrete data, we model the variables grand theft (GT), petty theft (PT), and extortion (E)
using generalized Poisson distributions. The variables violation of court order (VCO) and intimate partner
simple assault (IPSA) are modeled using negative binomial distributions. We obtain the true values of the
marginal distribution parameters $\boldsymbol{(\theta,\lambda,r,p)}$ by applying the method of moments to the dataset,
with $r$ rounded up to integers. To determine the specified correlation matrix, we estimate the empirical
Pearson correlation coefficients for the selected variables. Simulated data are generated with a sample size
of $N=1200$ across 1{,}000 replications to recover the target marginal distribution parameters and pairwise
correlation values observed in the LA crime dataset. The results are summarized in Table~11. The results show
that the estimated values are close to the true parameter values, with small relative bias (RB) values across
both marginal parameters and correlation estimates. In spite of a few standardized bias (SB) values being
relatively higher, the RMSE values remain low, and the coverage rates stay close to the nominal 95\% level.
Overall, the method demonstrates strong and stable performance in this mixed-distribution setting.


\begin{table}[H]
\centering
\caption{ Results from the LA Crime dataset replication study ($N = 1200$).}
\label{tab:LAcrime_replication}
\scriptsize
\setlength{\tabcolsep}{5pt}
\renewcommand{\arraystretch}{0.92}

\begin{tabular}{lrrrrrrr}
\toprule
      & TV     & AE     & SD     & RB     & SB     & RMSE   & CR (\%) \\
\midrule
\multicolumn{8}{l}{\textbf{Marginal parameters}} \\
$\theta_{1(\mathrm{GT})}$   & 3.5021 & 3.5041 & 0.0920 & 0.0555 & 2.1131  & 0.0919 & 95.3 \\
$\theta_{2(\mathrm{PT})}$   & 1.4921 & 1.4930 & 0.0475 & 0.0645 & 2.0260  & 0.0475 & 95.4 \\
$\theta_{3(\mathrm{E})}$    & 1.4864 & 1.4857 & 0.0461 & 0.0484 & 1.5605  & 0.0461 & 95.7 \\
$\lambda_{1(\mathrm{GT})}$  & 0.3519 & 0.3517 & 0.0156 & 0.0525 & 1.1860  & 0.0156 & 94.6 \\
$\lambda_{2(\mathrm{PT})}$  & 0.1590 & 0.1580 & 0.0198 & 0.5991 & 4.8106  & 0.0198 & 94.6 \\
$\lambda_{3(\mathrm{E})}$   & 0.1622 & 0.1622 & 0.0194 & 0.0390 & 0.3256  & 0.0194 & 94.1 \\
$r_{1(\mathrm{VCO})}$       & 4      & 4.0248 & 0.3133 & 0.6199 & 7.9145  & 0.3141 & 94.1 \\
$r_{2(\mathrm{IPSA})}$      & 7      & 7.0020 & 0.3601 & 0.0287 & 0.5575  & 0.3599 & 95.4 \\
$p_{1(\mathrm{VCO})}$       & 0.4069 & 0.4078 & 0.0194 & 0.2067 & 4.3462  & 0.0194 & 93.4 \\
$p_{2(\mathrm{IPSA})}$      & 0.1384 & 0.1383 & 0.0063 & 0.0734 & 1.6141  & 0.0063 & 94.6 \\
\midrule
\multicolumn{8}{l}{\textbf{Correlation}} \\
$\rho_{12}$ & 0.2291  & 0.2285 & 0.0289 & 0.2612 & 2.0730  & 0.0289 & 94.4 \\
$\rho_{13}$ & 0.1594  & 0.1573 & 0.0276 & 1.3397 & 7.7412  & 0.0277 & 95.3 \\
$\rho_{14}$ & 0.1124  & 0.1097 & 0.0289 & 2.3343 & 9.0729  & 0.0290 & 95.2 \\
$\rho_{15}$ & -0.0568 & -0.0560& 0.0277 & 1.2902 & 2.6471  & 0.0277 & 96.7 \\
$\rho_{23}$ & 0.1384  & 0.1383 & 0.0288 & 0.0824 & 0.3963  & 0.0288 & 94.1 \\
$\rho_{24}$ & 0.1013  & 0.1026 & 0.0291 & 1.2561 & 4.3787  & 0.0291 & 94.6 \\
$\rho_{25}$ & 0.1170  & 0.1132 & 0.0285 & 3.2296 & 13.2537 & 0.0287 & 94.4 \\
$\rho_{34}$ & 0.1749  & 0.1774 & 0.0298 & 1.4091 & 8.2611  & 0.0299 & 92.9 \\
$\rho_{35}$ & 0.1495  & 0.1509 & 0.0275 & 0.9319 & 5.0572  & 0.0276 & 94.8 \\
$\rho_{45}$ & 0.1698  & 0.1688 & 0.0288 & 0.5502 & 3.2471  & 0.0288 & 94.2 \\
\bottomrule
\end{tabular}
\end{table}

\section{Discussion}
We present a generalized methodology for generating multivariate discrete data with generalized Poisson, negative binomial, and binomial marginal distributions. Building on the framework of Demirtas (2006), our main contribution is a unified approach that preserves specified marginal distributional properties while accommodating a broad range of correlation structures and dimensional settings. As presented in Sections~3 and~4, the method performs well across both small and large sample scenarios. While not demonstrated in this paper, the proposed approach can be easily scaled to high-dimensional settings. Beyond the distributions studied in this paper, the framework is adaptable to other discrete distributions, such as geometric and hypergeometric distributions. The primary computational challenge of this approach lies in identifying suitable intermediate binary correlations, which requires extensive replications. However, once these correlations are obtained, generating data of any size is computationally plausible.

Our method has limitations. As mentioned in previous sections, the algorithm relies upon generating intermediate multivariate binary variables. The binary correlations are bounded by the lower and upper limits imposed by marginal expectations of the individual binary variates. When binary correlations fall outside these limits, the method becomes unreliable. Additionally, if splitting the discrete outcomes cannot yield binary variables with expectations close to $0.5$, the achievable range of correlation specifications becomes restricted. Overall, the proposed method offers a useful and flexible tool for researchers to simulate realistic, correlated discrete datasets with a straightforward implementation.

\section*{Funding statement}

The authors received no specific funding for this work.

\section*{COI statement}
The authors declare that they have no conflicts of interest for this work.

\section*{Disclosure statement}
The authors report there are no competing interests to declare.

\section*{References}

\begingroup
\setlength{\parindent}{0pt}            
\setlength{\parskip}{6pt}              
\everypar{\hangindent=2em\hangafter=1} 

Amatya, A., and Demirtas, H. (2015). Simultaneous generation of multivariate mixed data with Poisson and normal marginals. \textit{Journal of Statistical Computation and Simulation}, 85(15), 3129--3139. \url{https://doi.org/10.1080/00949655.2014.953534}

Bates, D., Maechler, M., Davis, T. A., Oehlschlagel, J., and Riedy, J. (2025). Sparse and Dense Matrix Classes and Methods. R package \textit{Matrix}. \url{https://cran.r-project.org/web/packages/Matrix/index.html}

Bonat, W. H. (2018). Multiple response variables regression models in R: The \textit{mcglm} package. \textit{Journal of Statistical Software}, 84(4), 1--30. \url{https://doi.org/10.18637/jss.v084.i04}

Consul, P. C. (1989). \textit{Generalized Poisson distributions: Properties and applications}. Decker, New York.

Deb, P., and Trivedi, P. K. (1997). Demand for medical care by the elderly: A finite mixture approach. \textit{Journal of Applied Econometrics}, 12(3), 313--336. \url{https://doi.org/10.1002/(SICI)1099-1255(199705)12:3<313::AID-JAE440>3.0.CO;2-G}

Demirtas, H. (2004). Simulation driven inferences for multiply imputed longitudinal datasets. \textit{Statistica Neerlandica}, 58(4), 466--482. \url{https://doi.org/10.1111/j.1467-9574.2004.00271.x}

Demirtas, H. (2006). A method for multivariate ordinal data generation given marginal distributions and correlations. \textit{Journal of Statistical Computation and Simulation}, 76(11), 1017--1025. \url{https://doi.org/10.1080/10629360600569246}

Demirtas, H. (2017). On accurate and precise generation of generalized Poisson variates. \textit{Communications in Statistics -- Simulation and Computation}, 46(1), 489--499. \url{https://doi.org/10.1080/03610918.2014.968725}

Demirtas, H., and Hedeker, D. (2011). A practical way for computing approximate lower and upper correlation bounds. \textit{The American Statistician}, 65(2), 104--109. \url{https://doi.org/10.1198/tast.2011.10090}

Emrich, L. J., and Piedmonte, M. R. (1991). A method for generating high-dimensional multivariate binary variates. \textit{The American Statistician}, 45(4), 302--304. \url{https://doi.org/10.1080/00031305.1991.10475828}

Fotouhi, A. R. (2008). Modelling overdispersion in longitudinal count data in clinical trials with application to epileptic data. \textit{Contemporary Clinical Trials}, 29(4), 547--554. \url{https://doi.org/10.1016/j.cct.2008.01.005}

Frechet, M. (1951). Sur les tableaux de corrélation dont les marges sont données. \textit{Annales de l’Université de Lyon, Section A}, 14, 53--77.

Higham, N. (2002). Computing the nearest correlation matrix -- A problem from finance. \textit{IMA Journal of Numerical Analysis}, 22(3), 329--343. \url{https://doi.org/10.1093/imanum/22.3.329}

Hoeffding, W. (1994). Scale--Invariant Correlation Theory. In Fisher, N. I., and Sen, P. K. (eds), \textit{The Collected Works of Wassily Hoeffding}. Springer Series in Statistics. Springer, New York, NY. \url{https://doi.org/10.1007/978-1-4612-0865-5_4}

Iovleff, S. (2019). \textit{MixAll: Clustering and classification using model-based mixture models}. R package. \url{https://CRAN.R-project.org/package=MixAll}

Li, H., Demirtas, H., and Chen, R. (2021). RNGforGPD: An R package for generation of univariate and multivariate generalized Poisson data. \textit{The R Journal}, 12(2), 173--188. \url{https://doi.org/10.32614/RJ-2021-007}

Madsen, L., and Dalthorp, D. (2007). Simulating correlated count data. \textit{Environmental and Ecological Statistics}, 14, 129--148. \url{https://doi.org/10.1007/s10651-007-0008-1}

Martin, I., Djuardi, Y., Sartono, E., Rosa, B. A., Supali, T., Mitreva, M., Houwing-Duistermaat, J. J., and Yazdanbakhsh, M. (2018). Dynamic changes in human-gut microbiome in relation to a placebo-controlled anthelminthic trial in Indonesia. \textit{PLOS Neglected Tropical Diseases}, 12(8), e0006620. \url{https://doi.org/10.1371/journal.pntd.0006620}

Minhajuddin, A. T., and Juarez, S. F. (2013). Simulating select families of multivariate discrete variates with positive correlations. \textit{WIREs Computational Statistics}, 5(5), 406--414. \url{https://doi.org/10.1002/wics.1271}

Soneson, C., and Delorenzi, M. (2013). A comparison of methods for differential expression analysis of RNA-Seq data. \textit{BMC Bioinformatics}, 14, 91. \url{https://doi.org/10.1186/1471-2105-14-91}

Thall, P. F., and Vail, S. C. (1990). Some Covariance Models for Longitudinal Count Data with Overdispersion. \textit{Biometrics}, 46(3), 657--671. \url{https://doi.org/10.2307/2532086}

Yahav, I., and Shmueli, G. (2012). On generating multivariate Poisson data in management science applications. \textit{Applied Stochastic Models in Business and Industry}, 28(1), 91--102. \url{https://doi.org/10.1002/asmb.901}

\endgroup

\end{document}